\documentclass[useAMS,usenatbib]{mn2e}
\usepackage{longtable,color}
\usepackage{graphicx}
\usepackage{mathrsfs}
\usepackage{epsfig}
\usepackage{natbib}
\usepackage{color}
\usepackage{float}
\usepackage{amsmath}
\usepackage{times}
\usepackage{upgreek}
\usepackage[varg]{txfonts}
\usepackage{enumerate}
\usepackage{verbatim}
\usepackage{booktabs}
\usepackage{multirow}
\usepackage{amssymb}
\usepackage{placeins}

\title[Bulge alignment]{Lack of Bulge Alignment in Late-type Galaxies with Large-scale Filaments Plausibly Unveils a Radial Migration Formation Scenario}
\author[Xue \& Rong]{Wenxiao Xue$^{1,2}$, Yu Rong$^{1,2}$\thanks{Corresponding author; E-mail: rongyua@ustc.edu.cn}\\
$^{1}$Department of Astronomy, University of Science and Technology of China, Hefei 230026, China\\
$^{2}$School of Astronomy and Space Sciences, University of Science and Technology of China, Hefei 230026, Anhui, China\\
}

\begin{document}
\maketitle

\begin{abstract}

The formation sequence of bulges and disks in late-type galaxies (LTGs) remains debated. While some theories suggest bulges form before disks, others propose the reverse. To address this, we analyze a bulge+disk decomposition catalog from the Sloan Digital Sky Survey, examining the alignment between central bulge major axes and large-scale filaments. For LTGs near filament spines, no significant alignment is found. However, LTGs farther from filaments show a marginal ($\sim 1\sigma$) perpendicular alignment. Notably, central bulges and outer disks exhibit strong alignment in the sky plane, suggesting bulge formation may be driven by material migration from outer disks. Although further simulations are needed, our results provide a novel alignment-based perspective on bulge formation, advancing understanding of galactic structural evolution.

\end{abstract}
\section{Introduction}
\label{sec:1}

In massive late-type galaxies (LTGs), a prominent central bulge is often present, with the outer regions primarily dominated by a stellar disk \citep[e.g.,][]{Simard11}. However, there is no consensus on the formation sequence of these components. One hypothesis suggests that the stellar disk forms first, followed by the bulge through gravitational disturbances or central instabilities or radial migrations \citep{Kormendy04,Guo11,Dalcanton97,Minchev12,Martig10}. Alternatively, some studies propose that a bulge forms initially within a dark matter halo, which later cools by accreting surrounding gas to form a stellar disk \citep{Immeli04a,Immeli04b,Carollo07}.

Observationally, the formation sequence of bulges and disks is often inferred by comparing the stellar population ages of these components. For example, \cite{Carollo07} used the colors of bulges and outer disks to estimate their stellar ages, though the use of color alone introduces significant uncertainties. Stellar color is influenced by both age and metallicity, making it challenging to distinguish the contributions of each. Furthermore, the presence of young stars—resulting from recent gas accretion and star formation—complicates age determination, as these stars contribute little to the total mass of the galaxy. Therefore, it is difficult to accurately assess the mass-weighted stellar population age from color alone, introducing uncertainty when inferring the formation sequence based on stellar population ages.

Moreover, even with accurate spectroscopic analysis of stellar populations, the ``inside-out'' formation mode of massive galaxies complicates the interpretation. This mode suggests that the inner regions of galaxies form stars first, followed by star formation in the outer regions \citep[e.g.,][]{Bai14, Kepner99, Tiret11, Schonrich17}. Consequently, even if a disk forms first, the central regions may form stars earlier than the outer disk, meaning the central bulge will also tend to be older than the outer disk. This overlap in age distribution makes it difficult to definitively resolve the formation sequence.

High-precision simulations, such as those from the FIRE project \citep{Ma20}, indicate that cold gas accretion in high-redshift galaxies can lead to the formation of massive stellar clumps that eventually form a bulge at the center of the disk \citep{Noguchi99,Noguchi18,Noguchi22,Kalita22,Elmegreen08,Immeli04a}. These results suggest that the bulge forms before the thin outer disk. However, large-scale cosmological simulations and semi-analytical models at lower redshifts indicate that most LTGs form a stellar disk first, followed by bulge formation due to disk instabilities \citep{Carollo07, Martig10}. Other studies propose that in some LTGs, a bar structure forms in the center rather than a bulge \citep{Combes93, Carollo99, Cameron10}. As a result, even with the aid of simulations, the formation sequence remains unclear.

To gain further insights, we turn to the alignment of galaxy structures with large-scale cosmic structures. Previous studies have shown that in filamentary structures, the stellar disks of LTGs weakly align with the direction of the filament due to the alignment of angular momentum with the filament's orientation \citep{Tempel13a, Tempel13b, Libeskind13,Rong16,Rong15b,Zhang15}. In contrast, early-type galaxies (ETGs) show a stronger alignment between their major axes and the filament direction, with their spin directions perpendicular to the filament \citep{Tempel13a, Tempel13b, Rong25}. This alignment is attributed to galaxy mergers, where the angular momentum of merger systems becomes perpendicular to the filament spine, leading to alignment of the major axes with the filament direction. Thus, studying the orientation of bulges and disks in LTGs within large-scale filamentary structures can offer valuable clues about their formation and evolution.

If the second hypothesis-that bulges form first due to galaxy mergers-is correct, we would expect the major axes of bulges to align with the filament spine. However, if the first hypothesis holds-where bulges form via instabilities/migration-the orientation of the bulge would be unrelated to the large-scale structure or even perpendicular to the filamentary spines. Therefore, examining the alignment of bulges in LTGs with their host filaments may shed light on the formation sequence of bulges and disks.

This study investigates the alignment of bulges in LTGs with their parent large-scale filaments. In section~\ref{sec:2}, we describe the sample selection. In section~\ref{sec:3}, we present the statistical analysis of bulge alignment with filaments. We summarize and discuss our findings in section~\ref{sec:4}.

\section{Sample}\label{sec:2}

We select LTGs from the spectroscopic sample of \cite{Simard11}, which provides comprehensive two-dimensional, point-spread-function-convolved bulge+disk decompositions for approximately 1.1 million galaxies in the Sloan Digital Sky Survey Data Release 7 \citep[SDSS DR7;][]{Abazajian09}. \cite{Simard11} implemented three distinct galaxy fitting models: a pure S\'ersic profile, a bulge+disk model with fixed S\'ersic index of bulge ($n_{\rm b} = 4$, and a bulge+disk model with free $n_{\rm b}$). For our analysis, we adopt the structural parameters derived from the free $n_{\rm b}$ bulge+disk model, whose robustness has been thoroughly validated in the original study \citep{Simard11}. The photometric uncertainties for both bulge and disk components remain below 0.1 magnitude for sources with bulge/disk magnitudes $g\simeq 19$~mag and $r\simeq 18.5$~mag.

The \cite{Simard11} catalog enables precise isolation of LTGs and distinct characterization of their central bulge and outer disk components, including position angles and ellipticities for each structural element. To ensure measurement precision, we restrict our sample to galaxies with bulge ellipticities $e > 0.2$. Furthermore, to exclude pseudobulges and bar structures, we impose a S\'ersic index threshold of $n_{\rm b} > 2.5$. for bulge components. We note that our results remain statistically consistent when employing alternative thresholds of $n_{\rm b}\geq 3$, $n_{\rm b}\geq 3.5$, or $n_{\rm b}\geq 4$.

Leveraging the morphology parameter $P_{pS}$ from the \cite{Simard11} catalog, we select LTGs based on the criterion $P_{pS} \leq 0.32$ \citep[see][for details]{Simard11}. Our analysis focuses on galaxies with substantial bulge components, specifically those with bulge-to-total flux ratios $B/T > 0.25$, ensuring minimal photometric contamination of the central bulge by the outer disk. This selection criterion also guarantees accurate position angle measurements for the bulge components. Concurrently, to investigate outer disk alignment, we exclude galaxies with excessively dominant bulges that might compromise disk photometry, thereby limiting our sample to the range $0.25 < B/T < 0.45$.

Stellar mass ($M_{\star}$) estimates for each LTG are derived from $r$-band magnitudes and $g-r$ colors using the mass-to-light ratio relation $\log(M_{\star}/L_r) = 1.097(g-r) - 0.306$ \citep{Bell03}.  To enhance the statistical significance of alignment signals, we restrict our analysis to LTGs with $M_{\star} > 10^{10} M_{\odot}$, as more massive galaxies typically exhibit stronger alignment signatures \citep{Tempel13a, Tempel15a}.

For each selected LTG, we identify the associated large-scale filament from the comprehensive catalog of \cite{Tempel14a}. The host filament is determined based on the three-dimensional spatial separation ($d_{\rm{gf}}$) between the galaxy and the filament spine. Following \cite{Wang24}, we focus on galaxies within $d_{\rm{gf}} \leq 1.0$ Mpc$/h$, as this distance effectively delineates the filament boundary. Our final sample comprises 12,106 LTGs with their respective host filaments for detailed analysis.


\section{Alignment of bulge}
\label{sec:3}

   \begin{figure*}
   \centering
   \includegraphics[width=0.8\textwidth, angle=0]{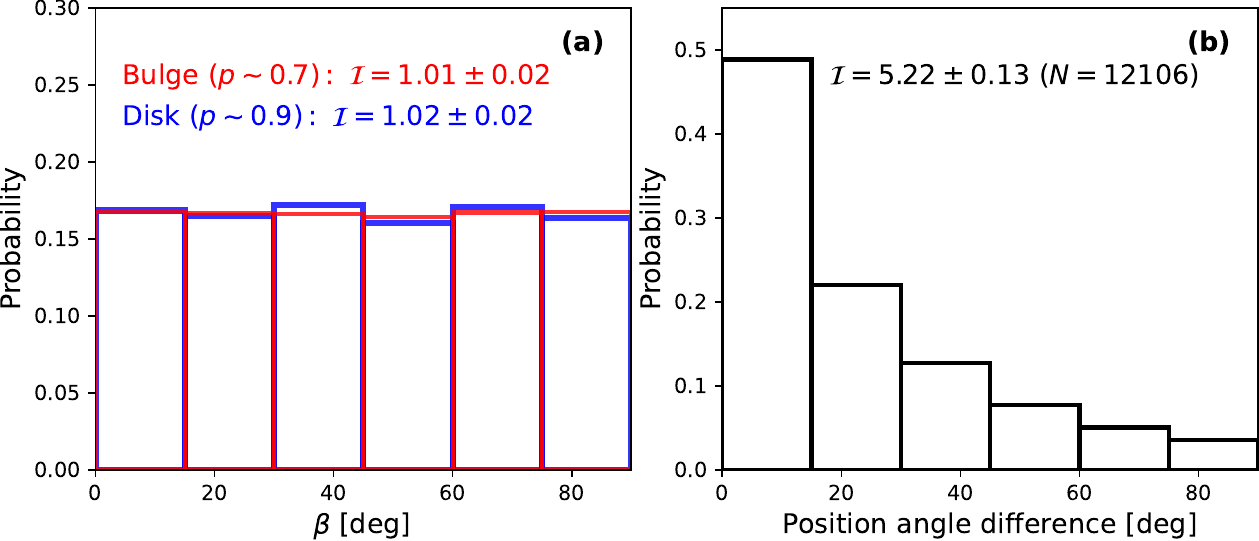}
   \caption{Panel a: Comparison of $\beta$ distributions for bulge (red) and disk (blue) components across 12,106 LTGs. High $p$-values from K-S tests indicate no alignment with filament orientation.
Panel b: Distribution of position angle differences between bulge and disk components in LTGs.}
   \label{fig1}
   \end{figure*}

  \begin{figure*}
   \centering
   \includegraphics[width=0.8\textwidth, angle=0]{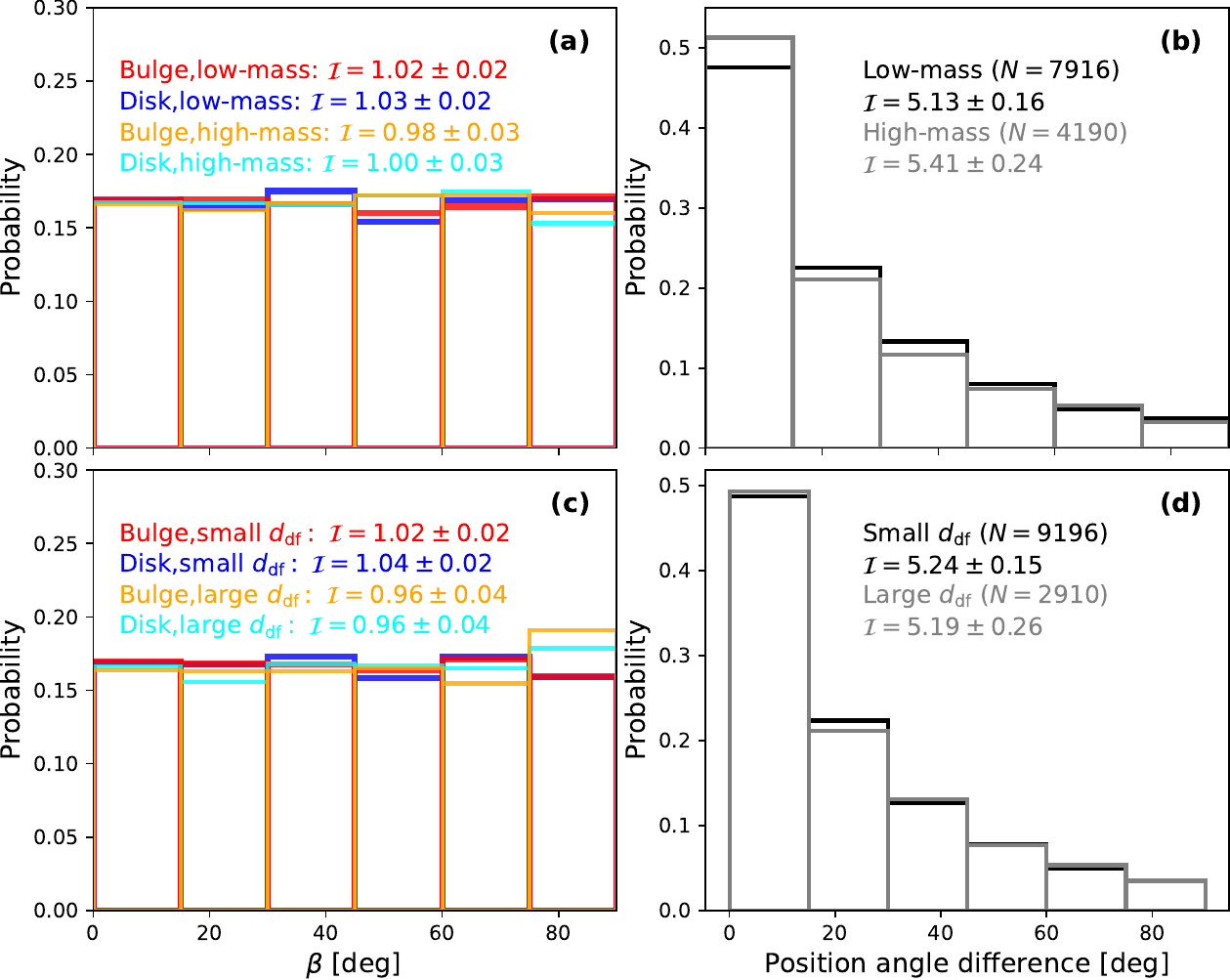}
   \caption{Panel a: Comparison of $\beta$ distributions for bulge and disk components in low-mass ($M_{\star}<10^{11}\ M_{\odot}$) and high-mass ($M_{\star}>10^{11}\ M_{\odot}$) LTGs. 
Panel b: Distributions of the position angle differences between the bulge and disk components in low-mass (black) and high-mass (gray) LTGs. Panel c: Comparison of $\beta$ distributions for bulge and disk components in LTGs located at distances $d_{\rm{gf}}<0.5$~Mpc$/h$ and $d_{\rm{gf}}>0.5$~Mpc$/h$ relative to filament spines.
Panel d: Distributions of position angle differences between bulge and disk components in LTGs with $d_{\rm{gf}}<0.5$~Mpc$/h$ (black) and $d_{\rm{gf}}>0.5$~Mpc$/h$ (gray).}
   \label{fig2}
   \end{figure*}

We measure the angle $\beta$ between the major axis and the orientation of the spine of the nearest filament for the bulge/disk components of each LTG on the celestial sphere, with $\beta$ restricted to the range [0,90$^\circ$]. An alignment signal is identified if the distribution of $\beta$ significantly deviates from a uniform distribution. Following the methodology of \cite{Rong24}, we define the parameter $I(\beta) = N_{0-45}/N_{45-90}$ to quantify the strength of the alignment signal, where $N_{0-45}$ and $N_{45-90}$ represent the number of galaxies with $\beta$ values in the ranges [0,45$^{\circ}$] and [45$^{\circ}$,90$^{\circ}$], respectively. A uniform distribution corresponds to $I(\beta) \approx 1$.

The error in ${\mathcal{I}}(\beta)$ is estimated using bootstrap resampling \citep{Rong25b}. From the original sample, we randomly select $N$ galaxies with replacement, repeating this 100 times to obtain 100 values of ${\mathcal{I}}(\beta)$, and the standard deviation of these values is taken as the uncertainty.

As shown in panel~a of Fig.\ref{fig1}, we observe no significant alignment between the major axes of the bulge and disk components of LTGs and their parent large-scale filaments, with ${\mathcal{I}}(\beta) = 1.01 \pm 0.02$ for the bulge component and ${\mathcal{I}}(\beta) = 1.02 \pm 0.02$ for the outer disk component. The K-S tests between the $\beta$ distributions and a uniform distribution yield large $p$-values ($p\sim 0.7$ and 0.9 for the bulge and disk components, respectively), further indicating no alignment signal.

However, we find a significant alignment between the position angles of the bulge and disk components, as illustrated in panel~b of Fig.\ref{fig1}. This alignment is not due to contamination between the two components during photometric measurements, as both components are sufficiently distinct. This suggests that the central bulges of LTGs may be plausibly influenced by or formed through the migration of material from the outer disks, resulting in an oblate spheroidal morphology.

Our analysis further reveals that the observed alignment patterns are independent of galactic masses, as demonstrated in panels~a and b of Fig.~\ref{fig2}. Both bulge and disk components in low-mass ($M_{\star}<10^{11}\ M_{\odot}$) and high-mass ($M_{\star}>10^{11}\ M_{\odot}$) LTGs consistently exhibit no statistically significant alignment with large-scale filaments, while maintaining mutual alignment within the sky plane. Nevertheless, the alignment signal appears to exhibit a potential dependence on the distance to the filament spine. As illustrated in panels~c and d of Fig.~\ref{fig2}, the structural components of LTGs located within $d_{\rm{gf}}<0.5$~Mpc$/h$ from the filament spine show no discernible alignment with their parent filaments. In contrast, LTGs positioned at relatively larger distances ($d_{\rm{gf}}>0.5$~Mpc$/h$) from filament spines manifest a potentially weak alignment signal (approximately at the 1$\sigma$ confidence level{\footnote{It is estimated as $|{\mathcal{I}}-1|/\sigma_{\mathcal{I}}$, where $\sigma_{\mathcal{I}}$ is the error of $\mathcal{I}$.}}), characterized by their major axes oriented perpendicular to the filament spines or their spin axes aligned parallel to the filament spines. The statistical significance of this alignment is further corroborated by the relatively small $p$-values ($p\sim0.01$) obtained from comparing the distributions of $\beta$ angles for both bulge and disk components in LTGs with $d_{\rm{gf}}>0.5$~Mpc$/h$ against a uniform distribution, thereby confirming the deviation of their orientations from isotropic distributions.

\section{Discussion}
\label{sec:4}

Using a comprehensive bulge+disk decomposition catalog derived from a large sample of late-type galaxies (LTGs), we investigate the alignment between the major axes of central bulge components and the orientations of their parent large-scale filaments. Our analysis reveals that for LTGs located in close proximity to filament spines ($d_{\rm{gf}}<0.5$~Mpc$/h$), no significant alignment signal is detected for the bulge components. In contrast, for LTGs situated at greater distances from filament spines ($d_{\rm{gf}}>0.5$~Mpc$/h$), the major axes of the bulges exhibit a weak tendency to be perpendicular to the filament spines, suggesting a subtle alignment between bulge spins and filament orientations. Notably, the major axes of both central bulges and outer disks demonstrate a strong mutual alignment on the sky plane. This observation plausibly implies that the central bulges in LTGs may be significantly influenced by, or even formed through, the migration of material from the outer disks, rather than through merger events.

It is important to emphasize that the proposed formation scenario for bulges in LTGs necessitates further investigation through hydrodynamical simulations. Recently, we employed the Illustris-TNG simulation \citep{Nelson19} to explore the formation of dynamically hot bulge structures in LTGs. Our preliminary findings indicate that disturbances and tidal heating induced by fly-by galaxies play a substantial role in bulge formation, rather than mergers (Wang et al. 2025, in preparation). Although the bulge formation scenario remains subject to further scrutiny and validation through simulations, our results provide a novel perspective on bulge formation mechanisms from an alignment standpoint, offering unique insights for future research endeavors.

We also acknowledge that previous studies have reported weak alignment signals between the disks/spins of LTGs and large-scale filaments, particularly for LTGs distant from filament spines, using deprojection methods \citep{Tempel13a,Tempel13b}. These findings are consistent with our observations, reinforcing the broader implications of our study.

\section*{Acknowledgments}

YR acknowledges supports from the CAS Pioneer Hundred Talents Program (Category B), NSFC grant 12273037, and the USTC Research Funds of the Double First-Class Initiative. This work is supported by the China Manned Space Program with grant no. CMS-CSST-2025-A06 and CMS-CSST-2025-A08.

\section*{Data Availability}

Data are available if requested.

\bibliographystyle{mn2e}

\begin{thebibliography}{}

\bibitem[\protect\citeauthoryear{Abazajian et al.}{2009}]{Abazajian09} Abazajian, K. et al. 2009, ApJS, 182, 543



\bibitem[\protect\citeauthoryear{Bai et al.}{2014}]{Bai14} Bai, L., et al. 2014, ApJ, 789, 134


\bibitem[\protect\citeauthoryear{Bell et al.}{2003}]{Bell03} Bell, E. F., McIntosh, D. H., Katz, N., Weinberg, M. D. 2003, ApJs, 149, 289

\bibitem[\protect\citeauthoryear{Cameron et al.}{2010}]{Cameron10} Cameron, E., Carollo, C. M., Oesch, P., et al. 2010, MNRAS, 409, 346

\bibitem[\protect\citeauthoryear{Carollo}{1999}]{Carollo99} 
Carollo, C. M., 1999, ApJ, 523, 566

\bibitem[\protect\citeauthoryear{Carollo et al.}{2007}]{Carollo07} Carollo, C. M., Scarlata, C., Stiavelli, M., Wyse, R. F. G., Mayer, L. 2007, ApJ, 658, 960





\bibitem[\protect\citeauthoryear{Combes \& Elmegreen}{1993}]{Combes93} Combes, F., Elmegreen, B. G. 1993, A\&A, 271, 391



\bibitem[\protect\citeauthoryear{Dalcanton et al.}{1997}]{Dalcanton97} Dalcanton, J. J., Spergel, D. N., Summers, F. J. 1997, ApJ, 482, 659







\bibitem[\protect\citeauthoryear{Elmegreen et al.}{2008}]{Elmegreen08} Elmegreen, B. G., Bournaud, F., Elmegreen, D. M. 2008, ApJ, 688, 67



\bibitem[\protect\citeauthoryear{Guo et al.}{2011}]{Guo11} Guo, Q., et al. 2011, MNRAS, 413, 101




\bibitem[\protect\citeauthoryear{Immeli et al.}{2004a}]{Immeli04a} Immeli, A., Samland, M., Gerhard, O., \& Westera, P. 2004a, A\&A, 413, 547

\bibitem[\protect\citeauthoryear{Immeli et al.}{2004b}]{Immeli04b} Immeli, A., Samland, M., Westera, P., \& Gerhard, O. 2004b, ApJ, 611, 20

\bibitem[\protect\citeauthoryear{Kalita et al.}{2022}]{Kalita22} Kalita, B. S., Daddi, E., Bournaud, F., et al. 2022, A\&A, 666A, 44

\bibitem[\protect\citeauthoryear{Kepner}{1999}]{Kepner99} Kepner, J. V. 1999, ApJ, 520, 59

\bibitem[\protect\citeauthoryear{Kormendy \& Kennicutt}{2004}]{Kormendy04} Kormendy, J., Kennicutt, R. C. J. 2004, Annual Review of Astronomy \& Astrophysics, 42, 603





\bibitem[\protect\citeauthoryear{Libeskind et al.}{2013}]{Libeskind13} Libeskind, N. I., Hoffman, Y., Steinmetz, M., Gottl\"ober, S., Knebe, A., Hess, S., 2013, ApJ, 766, L15


\bibitem[\protect\citeauthoryear{Ma et al.}{2020}]{Ma20} Ma X., et al., 2020, MNRAS, 493, 4315

\bibitem[\protect\citeauthoryear{Martig \& Bournaud}{2010}]{Martig10} Martig, M., Bournaud, F. 2010, ApJ Letters, 714L, 275


\bibitem[\protect\citeauthoryear{Minchev et al.}{2012}]{Minchev12} Minchev, I., Famaey, B., Quillen, A. C., Di Matteo, P., Combes, F., Vlaji\'c, M., Erwin, P., Bland-Hawthorn, J. 2012, A\&A, 548A, 126

\bibitem[\protect\citeauthoryear{Nelson et al.}{2019}]{Nelson19} Nelson, D., Springel, V., Pillepich, A., et al. 2019, Comput. Astrophys. Cosmol., 6, 2

\bibitem[\protect\citeauthoryear{Noguchi}{1999}]{Noguchi99} Noguchi, M. 1999, ApJ, 514, 77

\bibitem[\protect\citeauthoryear{Noguchi}{2018}]{Noguchi18} Noguchi, M. 2018, Nature, 559, 585

\bibitem[\protect\citeauthoryear{Noguchi}{2022}]{Noguchi22} Noguchi, M. 2022, MNRAS, 510, 1772









\bibitem[\protect\citeauthoryear{Rong et al.}{2024}]{Rong24} Rong, Y., Shen, J., Hua, Z. 2024, MNRAS, 531L, 9

\bibitem[\protect\citeauthoryear{Rong et al.}{2016}]{Rong16} Rong, Y., Liu, Y., Zhang, S.-N. 2016, MNRAS, 455, 2267




\bibitem[\protect\citeauthoryear{Rong \& Wang}{2025a}]{Rong25} Rong, Y., Wang, P. 2025, ApJ, 983, 122

\bibitem[\protect\citeauthoryear{Rong \& Wang}{2025b}]{Rong25b} Rong, Y., Wang, P., Tang, X.-X. 2025, ApJ Letters, 983, 3

\bibitem[\protect\citeauthoryear{Rong et al.}{2015}]{Rong15b} Rong, Y., Zhang, S.-N., Liao, J.-Y. 2015, MNRAS, 453, 1577


\bibitem[\protect\citeauthoryear{Sch\"onrich \& McMillan}{2017}]{Schonrich17} Sch\"onrich, R., McMillan, P. J. 2017, MNRAS, 467, 1154


\bibitem[\protect\citeauthoryear{Simard et al.}{2011}]{Simard11} Simard, L., Mendel, J. T., Patton, D. R., Ellison, S. L., McConnachie, A. W. 2011, ApJs, 196, 11



\bibitem[\protect\citeauthoryear{Tempel et al.}{2015}]{Tempel15a} Tempel, E., Guo, Q., Kipper, R., Libeskind, N. I. 2015, MNRAS, 450, 2727

\bibitem[\protect\citeauthoryear{Tempel \& Libeskind}{2013}]{Tempel13b} Tempel, E., Libeskind, N. I. 2013, ApJ, 775, L42

\bibitem[\protect\citeauthoryear{Tempel et al.}{2014}]{Tempel14a} Tempel, E., Stoica, R. S., Mart\'inez, V. J., Liivam\"agi, L. J., Castellan, G., Saar, E., 2014, MNRAS, 438, 3465

\bibitem[\protect\citeauthoryear{Tempel et al.}{2013}]{Tempel13a} Tempel, E., Stoica, R. S., Saar, E. 2013, MNRAS, 428, 1827




\bibitem[\protect\citeauthoryear{Tiret et al.}{2011}]{Tiret11} Tiret, O., Salucci, P., Bernardi, M., Maraston, C., Pforr, J. 2011, MNRAS, 411, 1435






\bibitem[\protect\citeauthoryear{Wang et al.}{2024}]{Wang24} Wang, W., Wang, P., Guo, H., et al. 2024, MNRAS, 532, 4604




\bibitem[\protect\citeauthoryear{Zhang et al.}{2015}]{Zhang15} Zhang, Y., Yang, X., Wang, H., et al. 2015, ApJ, 798, 17

\end{thebibliography}


\end{document}